\documentclass[reprint,aps,amsmath,amssymb,twoside,prd,showkeys,superscriptaddress]{revtex4-1}
\pdfoutput=1
\usepackage{verbatim}
\usepackage{comment}
\usepackage{graphicx}
\usepackage[T1]{fontenc}
\usepackage{epsfig}
\usepackage{bm}
\usepackage{amssymb}
\usepackage{float}
\usepackage{amsmath}
\usepackage{subfigure}
\usepackage{dcolumn}
\usepackage{cancel}
\usepackage[colorlinks]{hyperref}
\usepackage[usenames,dvipsnames]{color}
\hypersetup{
     breaklinks=true,
    pdfstartview={FitH},    
    colorlinks=true,       
    linkcolor=blue,          
    citecolor=red,        
    filecolor=magenta,      
    urlcolor=blue,           
    anchorcolor=green,      
    linktocpage=true
}

\renewcommand{\(}{\left(}

\def\d{{\rm d}}
\def\be{\begin{equation}}
\def\ee{\end{equation}}
\def\bea{\begin{eqnarray}}
\def\eea{\end{eqnarray}}

\newcommand{\HCd}{\mathcal{H}}

\def\HCdt0{\tilde{\HCd}_{0}}

\newcommand{\afffias}{Frankfurt Institute for Advanced Studies (FIAS), 
Ruth-Moufang-Strasse~1, 60438 Frankfurt am Main, Germany}
\newcommand{\affbgu}{Physics Department, Ben-Gurion University of the Negev, 
Beer-Sheva 
84105, Israel}
\begin{document}
\title{Quantifying the $S_8$ tension with the Redshift Space Distortion data set}
\author{David Benisty}
\email{benidav@post.bgu.ac.il}
\affiliation{\affbgu}\affiliation{\afffias}
\begin{abstract}
One problem of the $\Lambda$CDM model is the tension between the $S_8$ found in Cosmic Microwave Background (CMB) experiments and the smaller one obtained from large-scale observations in the late Universe. The $\sigma_8$ quantifies the relatively high level of clustering. Bayesian Analysis of the Redshift Space Distortion (RSD) selected data set yields: $S_8 = 0.700^{+0.038}_{-0.037}$. The fit has $3\sigma$ tension with the Planck 2018 results. With Gaussian processes method a model-independent reconstructions of the growth history of matter in-homogeneity is studied. The fit yields $S_8 = 0.707^{+0.085}_{-0.085}, 0.701^{+0.089}_{-0.089}$, and $ 0.731^{+0.063}_{-0.062}$ for different kernels. The tension reduces and being smaller then $1.5\, \sigma$. With future measurements the tension may be reduced, but the possibility the tension is real is a plausible situation.
\end{abstract}
\maketitle
%

\section{Introduction}
{One of latest breakthroughs in cosmology is the fact that our universe is not only expanding but also accelerating. This fact is proven from different data sets, such as Supernovae type Ia (SNIa) 
\cite{Schmidt:1998ys,Perlmutter:1999jt,Efstathiou:1998qr,Tonry:2003zg,Betoule:2014frx,Huang:2017zyu,Scolnic:2017caz,DiValentino:2020evt,Staicova:2016pfd}, cosmic chronometers \cite{Farooq:2016zwm,2018ApJ...861..126R,Riess:2018uxu,Nakar:2020pyd} and Baryon Acoustic Oscillation (BAO) \cite{Eisenstein:2005su,Reid:2009xm,Percival:2009xn,Kazin:2009cj,Blake:2011rj,Reid:2012sw,Alam:2016hwk,Abbott:2017wcz,Gil-Marin:2018cgo,Alam:2020sor} and the Cosmic Microwave Background (CMB) radiation analysis \cite{Melchiorri:1999br,deBernardis:2000sbo,Balbi:2000tg,Hinshaw:2012aka,Akrami:2018vks,Aghanim:2018eyx}. Assuming homogeneous and isotropic volume, the accelerated expansion is explained by the presence of the dark energy  \cite{RevModPhys.61.1,Copeland:2006wr,Amendola:2010ub,Amendola:2010bk,Mehrabi:2018dru,Mehrabi:2018oke,Ygael:2020lkn,Frieman:2008sn,Ade:2015xua}.}

The $\Lambda$CDM model suffers from well known problems \cite{Weinberg:2000yb,Peebles:2002gy}, such as 
the coincidence problem and the disagreement between the measured value of the vacuum energy density and the predicted one from Quantum Field Theory. Despite the good agreement with the majority of cosmological 
data \cite{Aghanim:2018eyx}, the model seems 
to be currently in tension with some recent measurements, such as the present value of the mass variance at 8$h^{-1}$Mpc,  namely the $\sigma_8$ tension \cite{Keeley:2019esp,Pandey:2019plg,Quelle:2019vam,Bhattacharyya:2018fwb,Lambiase:2018ows,Lin:2019htv,Berbig:2020wve,Quelle:2019vam}. There is $2\sigma$ tension between the constraints from Planck on the matter density $\Omega_m^{(0)}$ and the amplitude $\sigma_8$ of matter fluctuations in linear theory and those from local measurements. {Planck derives $\sigma_8 = 0.811 \pm 0.006$ \cite{Aghanim:2018eyx}, local measurements find smaller values:  $0.75 \pm 0.03$ from Sunyaev-Zeldovich cluster counts \cite{Ade:2013lmv}, $0.808^{+0.009}_{-0.017}$ from DES \cite{Abbott:2017wau} and $ 0.772\pm0.029$ from KiDS-450 weak-lensing surveys \cite{Hildebrandt:2016iqg}. Another parameter that quantifies the matter fluctuations is the combination:}
\begin{equation}
S_8 = \sigma_8 \sqrt{\Omega_{m,0}/0.3}.
\end{equation}
{where $\Omega_{m,0}$ is the matter density.
There are many claims how to solve the tension: from the observational point of view or from a theoretical point of view \cite{Wang:2002rta,Maeder:2017ksf,Hoyle:2010ce,Meerburg:2014bpa,Moresco:2017hwt,Anagnostopoulos:2017iao,Kazantzidis:2018rnb,Gannouji:2018ncm,Kazantzidis:2018jtb,Perivolaropoulos:2019vkb,Wang:2019ufm,Kazantzidis:2019dvk,Kazantzidis:2020tko,Alestas:2020mvb,Benisty:2020nql,Barros:2018efl,Anagnostopoulos:2019miu,DiValentino:2019jae,DiValentino:2019ffd,Vagnozzi:2019ezj,Akarsu:2020pka,Benisty:2020vvm,Benisty:2020nql,Benisty:2020nuu,Benisty:2019pxb,Banerjee:2019kgu,Anagnostopoulos:2019myt,Benisty:2018qed,Benisty:2017eqh,Benisty:2017rbw,Benisty:2017lmt,Fay:2020egt,Espinosa:2020qtq,Geng:2020mga,Sola:2019lnw,Basilakos:2019zsf,Marinucci:2020weg}. Here we use the updated data-sets of the $f\sigma_8 $ measurements, including the collected data set from 2006-2018 \cite{Blake:2012pj,Jones:2004zy,Alam:2015mbd,Wang:2017wia,Guzzo:2013spa} (collected by \cite{Kazantzidis:2018rnb}) and the completed SDSS extended eBOSS Survey, DES and others \cite{deMattia:2020fkb,Tamone:2020qrl,Aubert:2020lfr,Zhao:2020tis,Gil-Marin:2020bct,Neveux:2020voa,Bautista:2020ahg,Said:2020epb,Qin:2019axr,Blake:2018tou,Zarrouk:2018vwy,Zhao:2018gvb,Percival:2018twa,Adams:2017val,Li:2016bis,Chuang:2016uuz,Sanchez:2016sas,Chuang:2016uuz,Marin:2015ula,Wang:2014qoa,Satpathy:2016tct,Okumura:2015lvp}. From more then 100 points we select different points from different redshifts. With Bayesian analysis and with Gaussian Process (GP) we analyse the $S_8$. }

The plan of the work is the following: Section \ref{sec:GMP} formulates the theoretical background for the standard models in cosmology. Section \ref{sec:BA} constraints the parameters with Likelihood analyses. Section \ref{sec:GPM} uses the Gaussian Process Regression method and estimates the corresponding cosmological parameters. Finally, section \ref{sec:Dis} summarizes the results.

\begin{figure}[t!]
 	\centering
\includegraphics[width=0.47\textwidth]{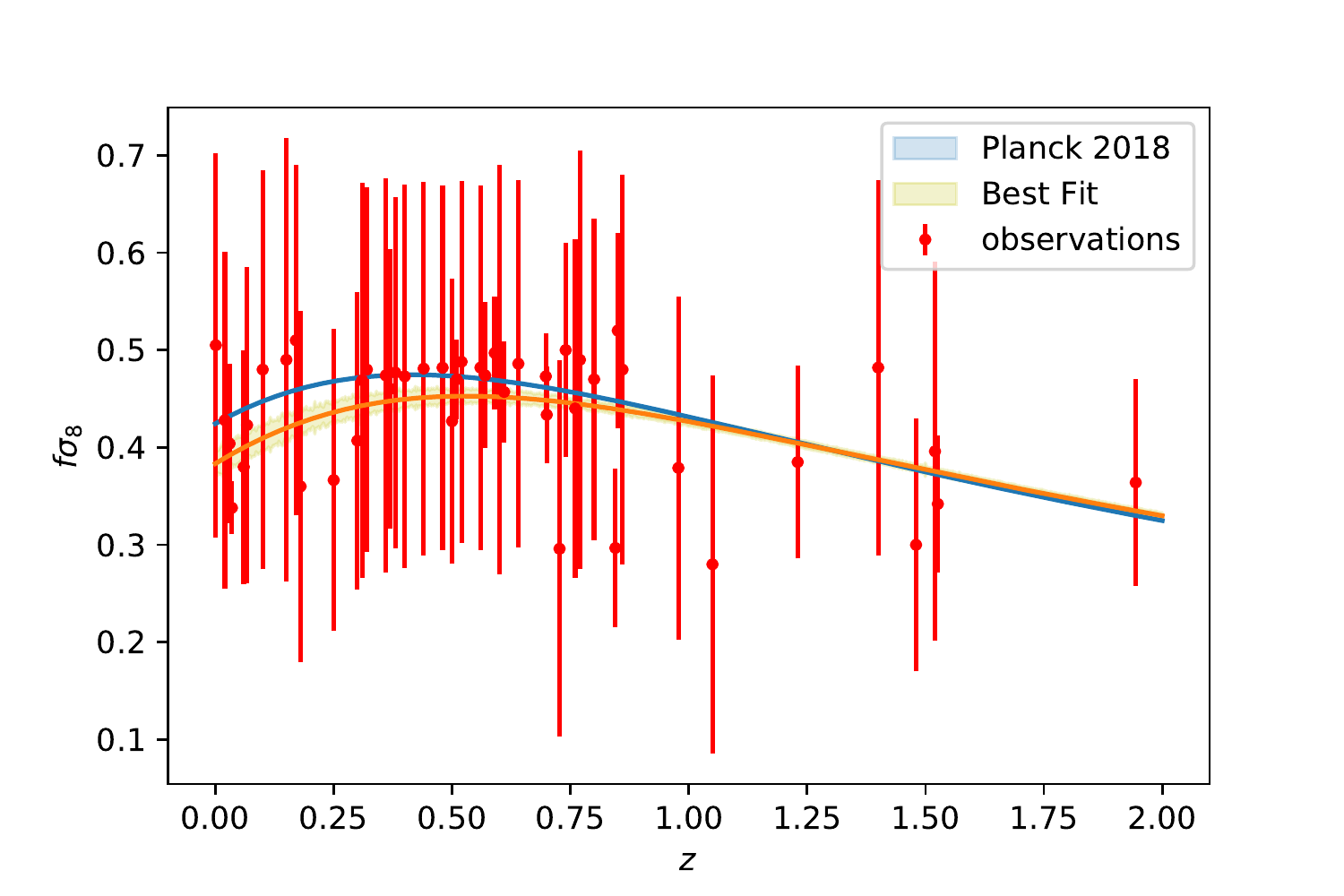}
\caption{\it{Growth of matter data set. The blue line is the prediction of Planck 2018 results with $5\sigma$ error. The yellow line preforms the best fit model with the full data set ($5\sigma$ error).}}
 	\label{fig:1}
\end{figure}

\begin{table}[t!]
\label{tab:data}
\begin{tabular}{ccccc}

Index & Survey & $z$ & $f\sigma_8(z)$ & Refs \\
\hline \hline
1 & 2MTF & 0.001 & $0.505 \pm 0.085$ &  \cite{Howlett:2017asq} \\
2 & 6dFGS+SnIa & $0.02$ & $0.428\pm 0.0465$ & \cite{Huterer:2016uyq}  \\
3 & 2MTF, 6dFGS & $0.03$ & $ 0.404^{+0.082}_{-0.081}$ & \cite{Qin:2019axr}\\
4 & 6dFGS, SDSS  & $0.035$ & $ 0.338 \pm 0.027$ & \cite{Said:2020epb}\\
5 & 6dFGS & $0.06$ & $ 0.38 \pm 0.12$ & \cite{Blake:2018tou}\\
6 & 6dFGS& $0.067$ & $0.423\pm 0.055$ & \cite{Beutler:2012px}\\
7 & SDSS DR13  & $0.1$ & $0.48 \pm 0.16$ & \cite{Feix:2016qhh} \\
8 & SDSS-MGS & $0.15$ & $0.490\pm0.145$ & \cite{Howlett:2014opa}\\
9 & 2dFGRS & $0.17$ & $0.510\pm 0.060$ & \cite{Song:2008qt}\\
10 & GAMA & $0.18$ & $0.360\pm 0.090$ & \cite{Blake:2013nif}\\
11 & SDSS-LRG-60 & $0.25$ & $0.3665\pm0.0601$ & \cite{Samushia:2011cs} \\
12 & SDSS-BOSS& $0.30$ & $0.407\pm 0.055$ & \cite{Tojeiro:2012rp} \\
13 & BOSS DR12 & $0.31$ & $0.469 \pm 0.098$ &  \cite{Wang:2017wia} \\
14 & SDSS DR10, DR11 & $0.32$ & $0.48 \pm 0.10$ & \cite{Sanchez:2013tga} \\
15 & BOSS DR12 & $0.36$ & $0.474 \pm 0.097$ &  \cite{Wang:2017wia} \\
16 & SDSS-LRG-200 & $0.37$ & $0.4602\pm 0.0378$ & \cite{Samushia:2011cs} \\
17 & BOSS DR12 & $0.38$ & $0.477 \pm 0.051$ & \cite{Beutler:2016arn} \\
18 & BOSS DR12 & $0.40$ & $0.473 \pm 0.086$ &  \cite{Wang:2017wia}\\
19 & BOSS DR12 & $0.44$ & $0.481 \pm 0.076$ &  \cite{Wang:2017wia}\\
20 & BOSS DR12 & $0.48$ & $0.482 \pm 0.067$ &  \cite{Wang:2017wia}\\
21 & SDSS-BOSS& $0.50$ & $0.427\pm 0.043$ & \cite{Tojeiro:2012rp} \\
22 & SDSS-III BOSS & $0.51  $ & $ 0.470 \pm 0.041  $& \cite{Sanchez:2016sas}\\
23 & BOSS DR12 & $0.52$ & $0.488 \pm 0.065$ &  \cite{Wang:2017wia} \\
24 & BOSS DR12 & $0.56$ & $0.482 \pm 0.067$ &  \cite{Wang:2017wia} \\
25 & BOSS 11 CMASS & $0.57$ & $ 0.438\pm0.037 $& \cite{Li:2016bis}\\
26 & SDSS-III DR12 & $0.59 $ & $ 0.497 \pm 0.058 $& \cite{Chuang:2016uuz}\\
27 & Vipers  & $0.6$ & $0.48 \pm 0.12$ & \cite{delaTorre:2016rxm} \\
28 & SDSS-III BOSS & $0.61 $ & $ 0.457 \pm 0.052 $& \cite{Chuang:2016uuz}\\
29 & BOSS DR12 & $0.64$ & $0.486 \pm 0.070$ &  \cite{Wang:2017wia}  \\
30 & SDSS-IV eBOSS & $0.698$ & $ 0.473 \pm 0.044$ & \cite{Gil-Marin:2020bct}\\
31 & eBOSS DR16 LRGxELG & $0.70$ & $0.4336\pm 0.05003$ & \cite{Zhao:2020tis}\\
32 & Vipers  & $0.727$ & $0.296 \pm 0.0765$ & \cite{Hawken:2016qcy}\\
33 & SDSS-IV eBOSS & $0.74$ & $ 0.50\pm0.11$ & \cite{Aubert:2020lfr}\\
34 &Vipers v7& $0.76$ & $0.440\pm 0.040$ & \cite{Wilson:2016ggz}\\
35 & VVDS & $0.77$ & $0.490\pm 0.18$ & \cite{Song:2008qt} \\
36 & Vipers& $0.80$ & $0.470\pm 0.080$ & \cite{delaTorre:2013rpa}\\
37 & SDSS-IV eBOSS & $0.845$ & $0.289^{+0.085}_{-0.096}$ & \cite{deMattia:2020fkb}\\
38 &  SDSS-IV eBOSS & $0.85$ & $0.35 \pm 0.10$ & \cite{Tamone:2020qrl}\\
39 & Vipers  & $0.86$ & $0.48 \pm 0.10$ & \cite{delaTorre:2016rxm} \\
40 & SDSS-IV eBOSS & $0.978$ & $ 0.379 \pm 0.176$ & \cite{Zhao:2018gvb} \\
41 &Vipers v7 & $1.05$ & $0.280\pm 0.080$ & \cite{Wilson:2016ggz} \\
42 & SDSS-IV eBOSS & $1.230 $ & $ 0.385 \pm 0.099$ & \cite{Zhao:2018gvb}\\
43 & FMOS & $1.4$ & $ 0.494^{+0.126}_{-0.120}$& \cite{Okumura:2015lvp}\\
44 & SDSS-IV eBOSS & $1.480$ & $0.439 \pm 0.048$ & \cite{Neveux:2020voa}\\
45 & SDSS-IV eBOSS & $1.52$ & $ 0.426 \pm 0.0777$ & \cite{Zarrouk:2018vwy}\\
46 & SDSS-IV eBOSS & $1.526$ & $ 0.342 \pm 0.070$ & \cite{Zhao:2018gvb}\\
47 & SDSS-IV eBOSS & $1.944$ & $ 0.364 \pm 0.106$ & \cite{Zhao:2018gvb}\\
\hline \hline
\end{tabular}
\caption{The selected Red Shift Space Distortion data set that be used in this paper.}
\end{table}

\section{Growth of matter perturbations}
\label{sec:GMP}
{In this section we present the basic equations required in our analysis. We begin with the Hubble parameter in a flat $\Lambda$CDM universe given by:}
\begin{equation}\label{eq:hubble-parameter}
	H(a)^2=H_0^2 \left[\Omega_{m,0}a^{-3}+(1-\Omega_{m,0}) a^{-3(1+w) }\right] \;,
\end{equation}
{where $H_0$ is the Hubble constant, $\Omega_{m,0}$ is the present day value of the matter density parameter, $a$ is the scale factor and $w$ is the dark energy equation of state. The scale factor is connected to the redshift via the relation $a = 1/(z+1)$. Since we will discuss on low redshifts ($z < 2$) we ignore the contribution of the radiation. The matter density can then also be expressed as a function of the scale factor:}
\begin{equation}\label{matter_density}
	\Omega_m(a)=\frac{\Omega_{m,0}a^{-3}}{H(a)^2/H^2_0}\;.
\end{equation}
{The matter density perturbations in Fourier space depend on the underlying cosmological model. The linear matter perturbations grow according to:}
\begin{equation}
\delta''_m(a)+\left(\frac{3}{a}+\frac{H'(a)}{H(a)}\right)\delta'_m(a)=\frac{3}{2}\frac{\Omega_m(a)}{a^2}\delta_m(a) \;.
\end{equation}
{The equation above has an analytical solution for the growing mode, given by \cite{Silveira:1994yq,Percival:2005vm,Belloso:2011ms,Nesseris:2015fqa}}
\begin{equation}
\frac{\delta_m(a)}{a}=  _2F_1\left(-\frac{1}{3w}\;,\frac{w-1}{2w}\;;1-\frac{5}{6w}\;;a^{-3w}(1-\frac{1}{\Omega_{m,0}})\right) \;.
\label{eq:delta}
\end{equation}
{The dependence on the wave number $k$ disappears because of the assumption of small scales approximation. We define the growth rate $f$ and the root mean square normalization of the matter power spectrum $\sigma_8$ as:}
\bea
f(a)&=&\frac{\d\log{\delta_m}}{\d\log{a}},\\
\sigma_8(a)&=&\sigma_{8,0}\frac{\delta_m(a)}{\delta_m(1)}\;,
\label{eq:fsig8z}
\eea
{A more robust and reliable quantity that is measured by redshift surveys is the combination of the growth rate $f(a)$ and $\sigma_8(a)$:}
\begin{equation}\label{eq:fs8-parameter}
f\sigma_8(a)=a\frac{\delta_m'(a)}{\delta_m(1)}\sigma_{8,0}\;.
\end{equation}
{Equation \eqref{eq:fs8-parameter} will be our key quantity, which will be tested with the most recent data available in the following sections. Fig. \ref{fig:1} shows the data set we use in this work that table \ref{tab:data} summarizes.}
\begin{figure*}[t!]
 	\centering
 		\label{fig:2}
\includegraphics[width=0.47\textwidth]{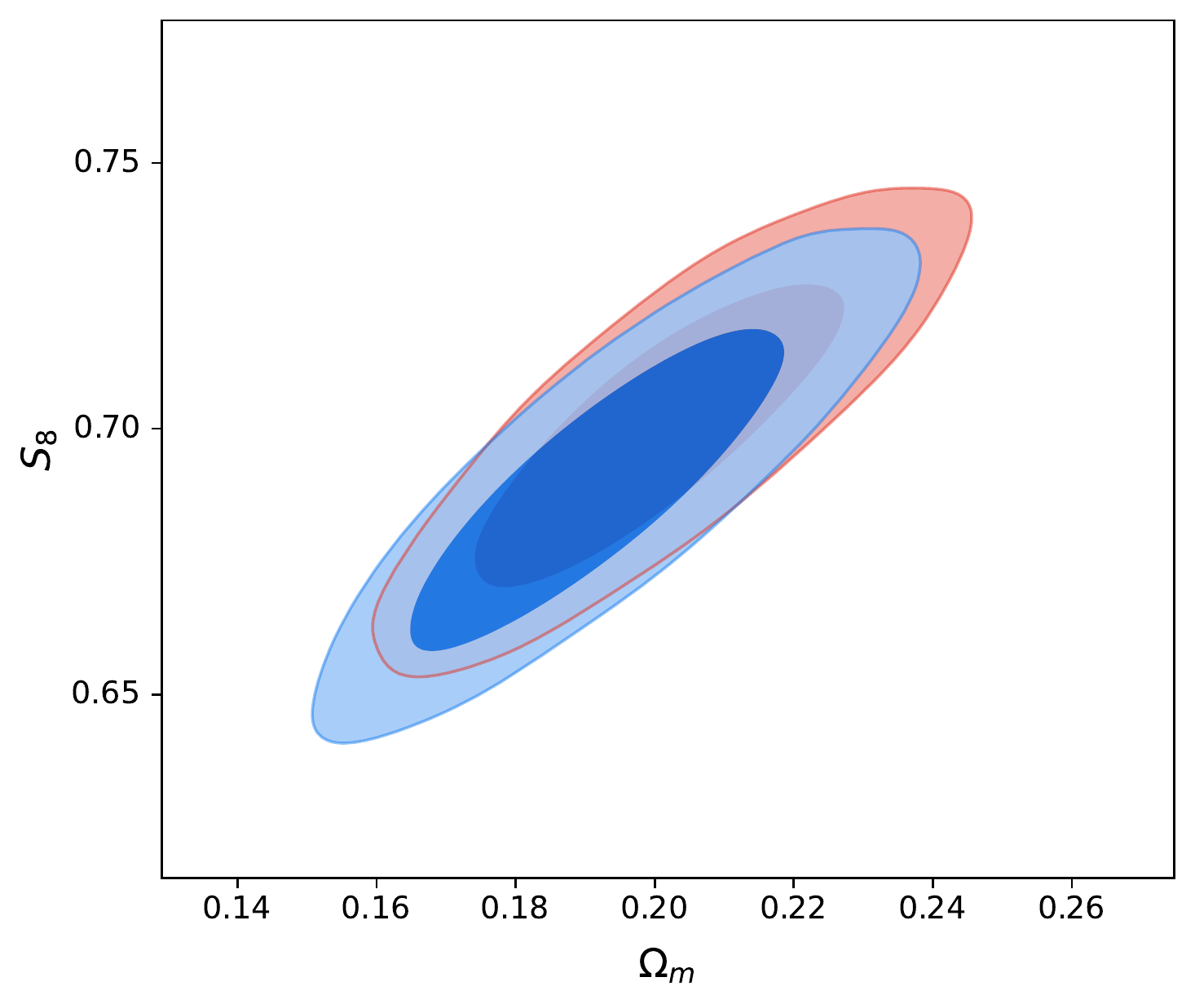}
\includegraphics[width=0.47\textwidth]{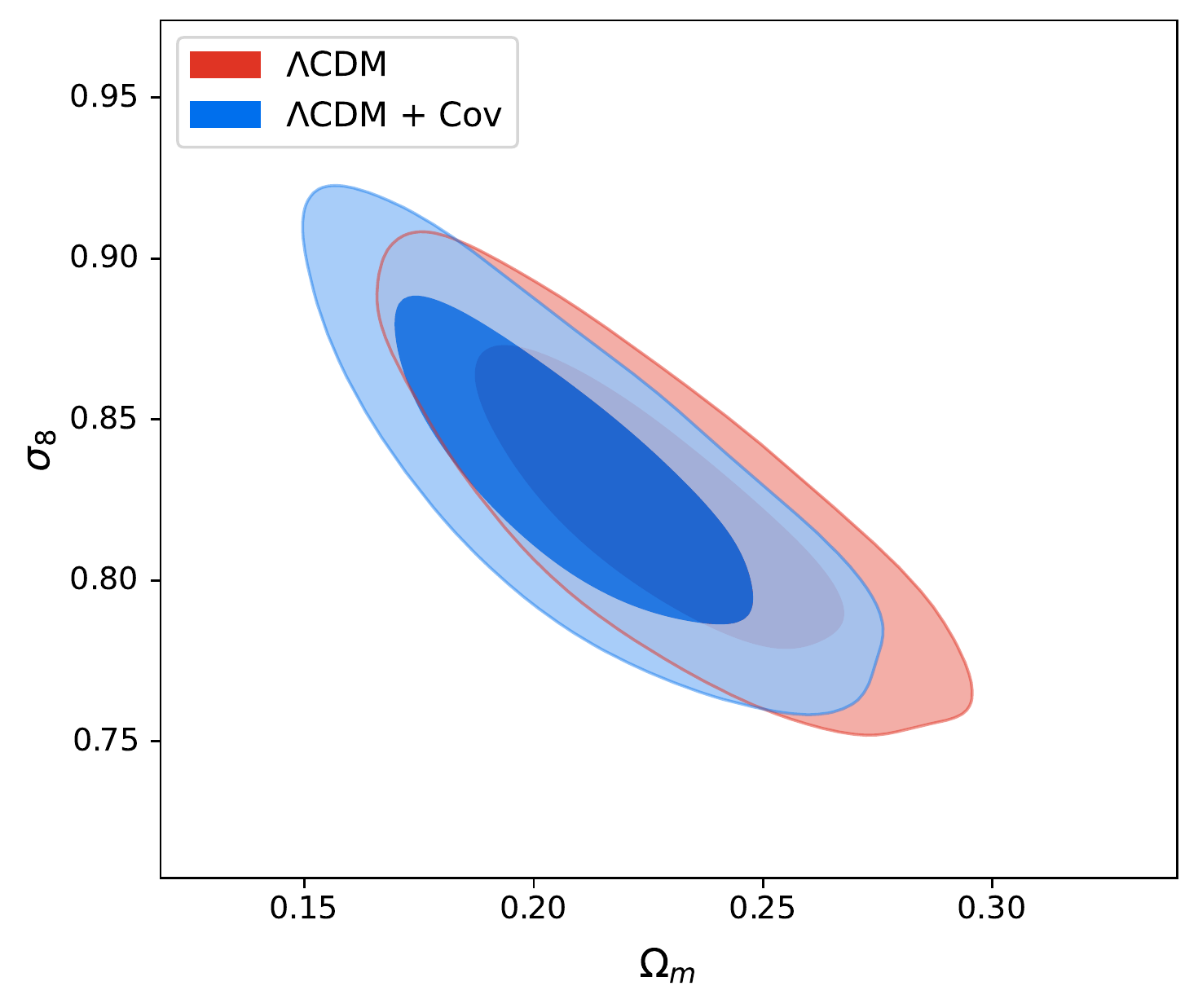}
\\
\includegraphics[width=0.47\textwidth]{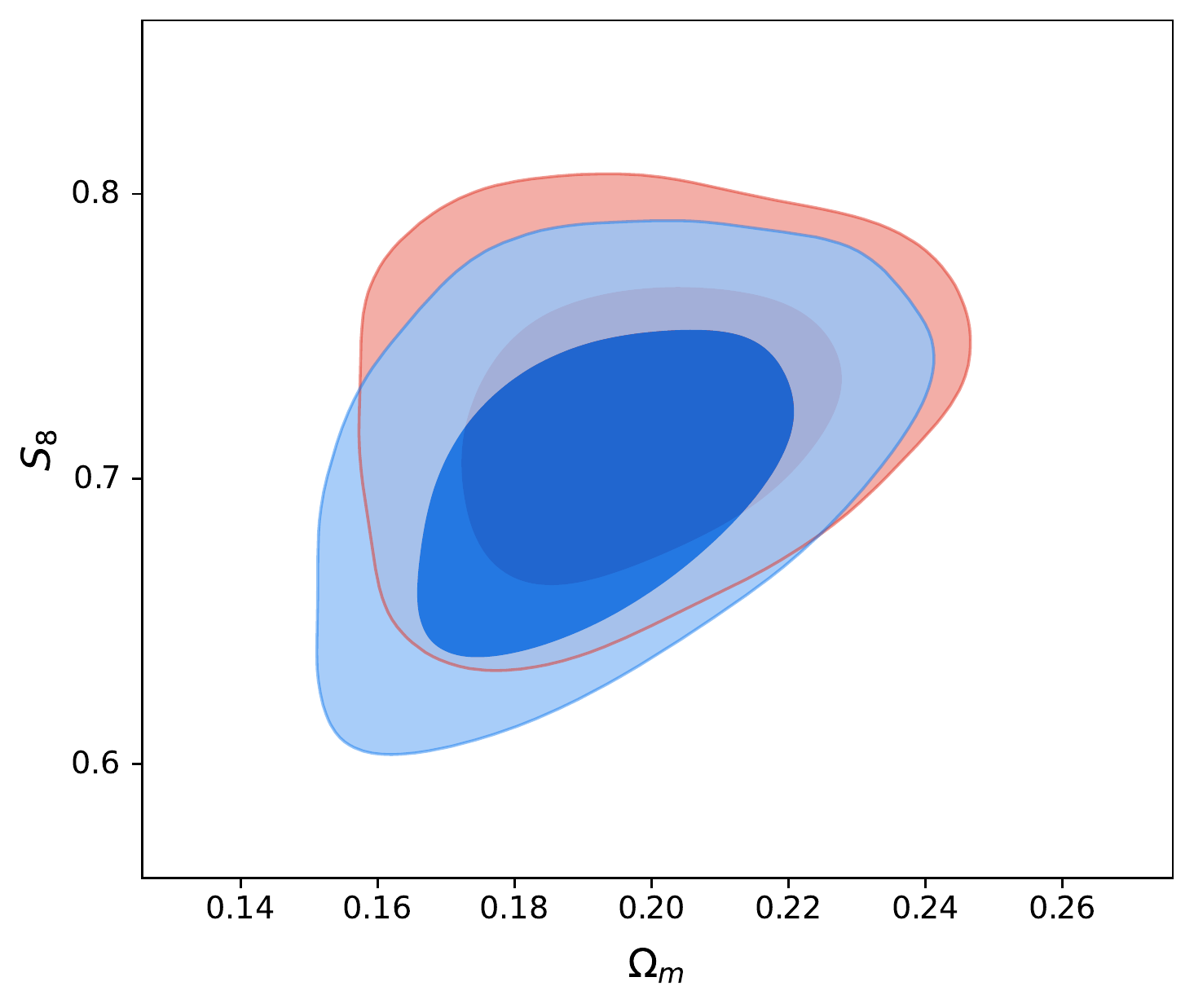}
\includegraphics[width=0.47\textwidth]{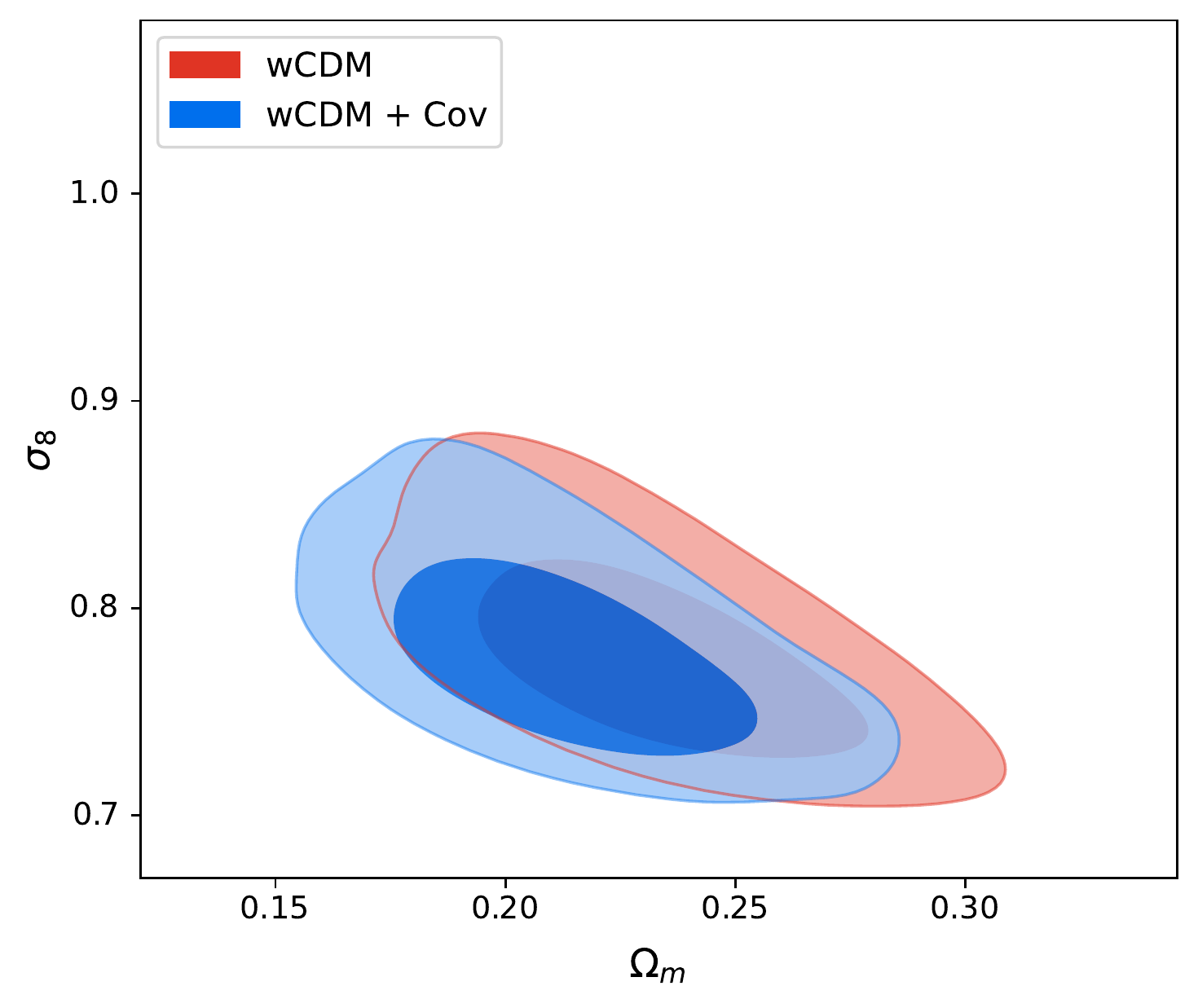}

\begin{tabular}{ccccc} \hline \hline
 Model & $\Omega_m$ & $\sigma_8$ & $S_8$ & w
 \vspace{0.05cm}\\ \hline
\hline
$\Lambda$CDM & $0.201^{+0.036}_{-0.033}   $ & $0.857^{+0.044}_{-0.042}   $& $0.700^{+0.038}_{-0.037}   $  & $-1$\\
$\Lambda$CDM + Ran Cov &  $0.192^{+0.037}_{-0.033}   $ &  $0.863^{+0.046}_{-0.043}   $&  $0.690^{+0.040}_{-0.038}   $    & $-1$\\
\hline\hline
wCDM &  $0.198^{+0.038}_{-0.035}   $ & $0.891^{+0.11}_{-0.095}$& $0.722^{+0.072}_{-0.067}$    & $-0.90^{+0.29}_{-0.30} $\\ 
wCDM + Ran Cov&  $0.192^{+0.038}_{-0.034}   $ & $0.879^{+0.10}_{-0.088} $& $0.701^{+0.077}_{-0.074}   $ & $-0.96^{+0.30}_{-0.31} $\\
\hline\hline
Planck 2018 & $0.311 \pm 0.019$ & $0.811 \pm  0.006$ & $0.834 \pm 0.016$ &$-1$\\ 
\hline\hline

\end{tabular}

\caption{\it{The posterior distribution for $\Lambda$CDM and wCDM. Rhe red distribution shows the fit for diagonal covariance and the blue distribution show the fit with random covariance between the point. The upper panel  presents the fit for $\Lambda$CDM and the lower panel shows the fit for wCDM. }}

\end{figure*}

\section{Bayesian Analysis}
\label{sec:BA}
{In order to test the standard models with the data sets, we use Bayesian Analysis. The $\chi^2$ between the models and the set is defined as: }
\begin{eqnarray}
\chi^2=V^i C_{ij}^{-1} V^j  \label{chi2eq}
\end{eqnarray}
{where $V^i= f\sigma _{8,i} - f\sigma _8 (z_i; \Omega_m,w,\sigma_8)$. Here $f\sigma _{8,i}$ corresponds to each of the data points. $f\sigma _8 (z_i; \Omega_m,w,\sigma_8)$ is the theoretical value for a given set of parameters values. The covariance matrix which has been assumed to leave the data points uncorrelated, with:}
\begin{equation}
C_{ii} = \sigma_i^2.
\end{equation}
{The effects of possible correlations among data points can be estimated by introducing a number of randomly selected nondiagonal elements in the covariance matrix while keeping it symmetric. In this approach we introduce positive correlations in 9 randomly selected pairs of data points (about $20\%$ of the data). The positions of the non-diagonal elements are chosen randomly and the magnitude of the randomly selected covariance matrix element $C_{ij}$ is set to}
\begin{equation} 
C_{ij} =0.5 \sigma_i \sigma_j
\label{cij}
\end{equation}
{where $\sigma_i \sigma_j$ are the published $1\sigma$ errors of the data points $i,j$. The prior is a uniform distribution, where $\Omega_{m} \in [0.01;0.9]$, $w\in[-2.5;-0.5]$ and $\sigma_8\in [0.5;1.2]$. Regarding the problem of data fit, we use the open-source sampler \textbf{emcee} \cite{Foreman_Mackey_2013} with the \textbf{GetDist} \cite{Lewis:2019xzd} to present the data.}
\begin{figure}
 	\centering
\includegraphics[width=0.45\textwidth]{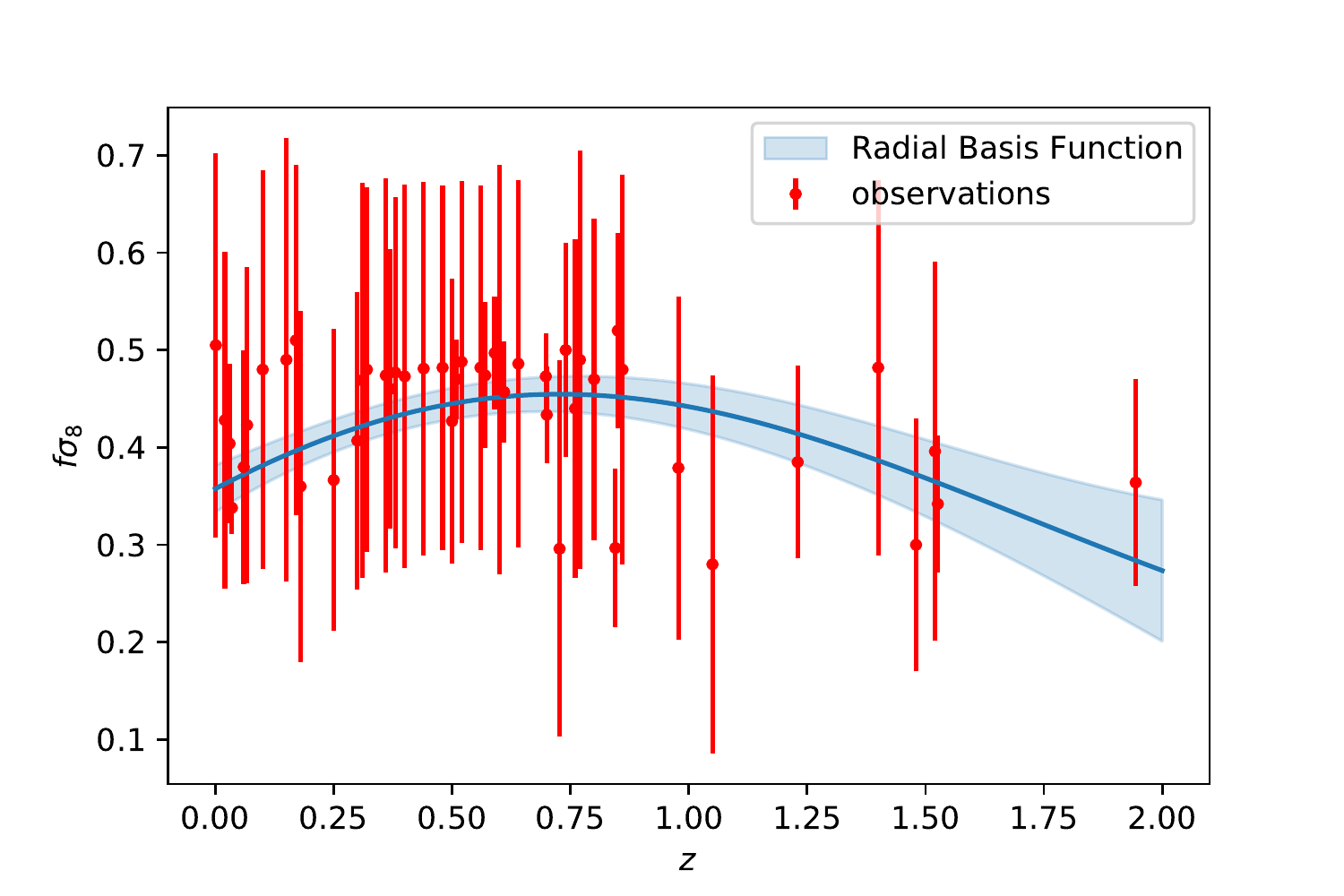}\\
\includegraphics[width=0.45\textwidth]{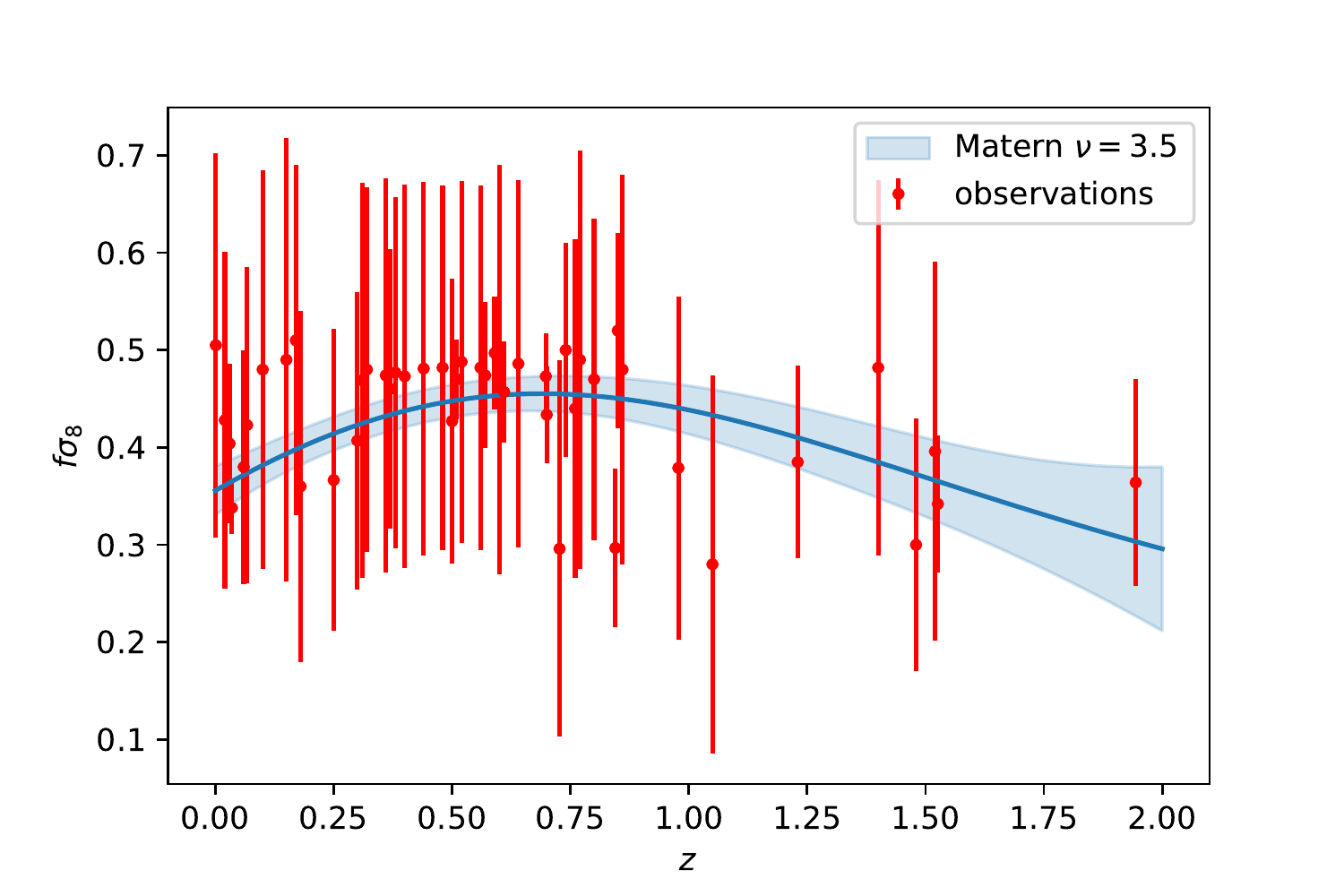}\\
\includegraphics[width=0.45\textwidth]{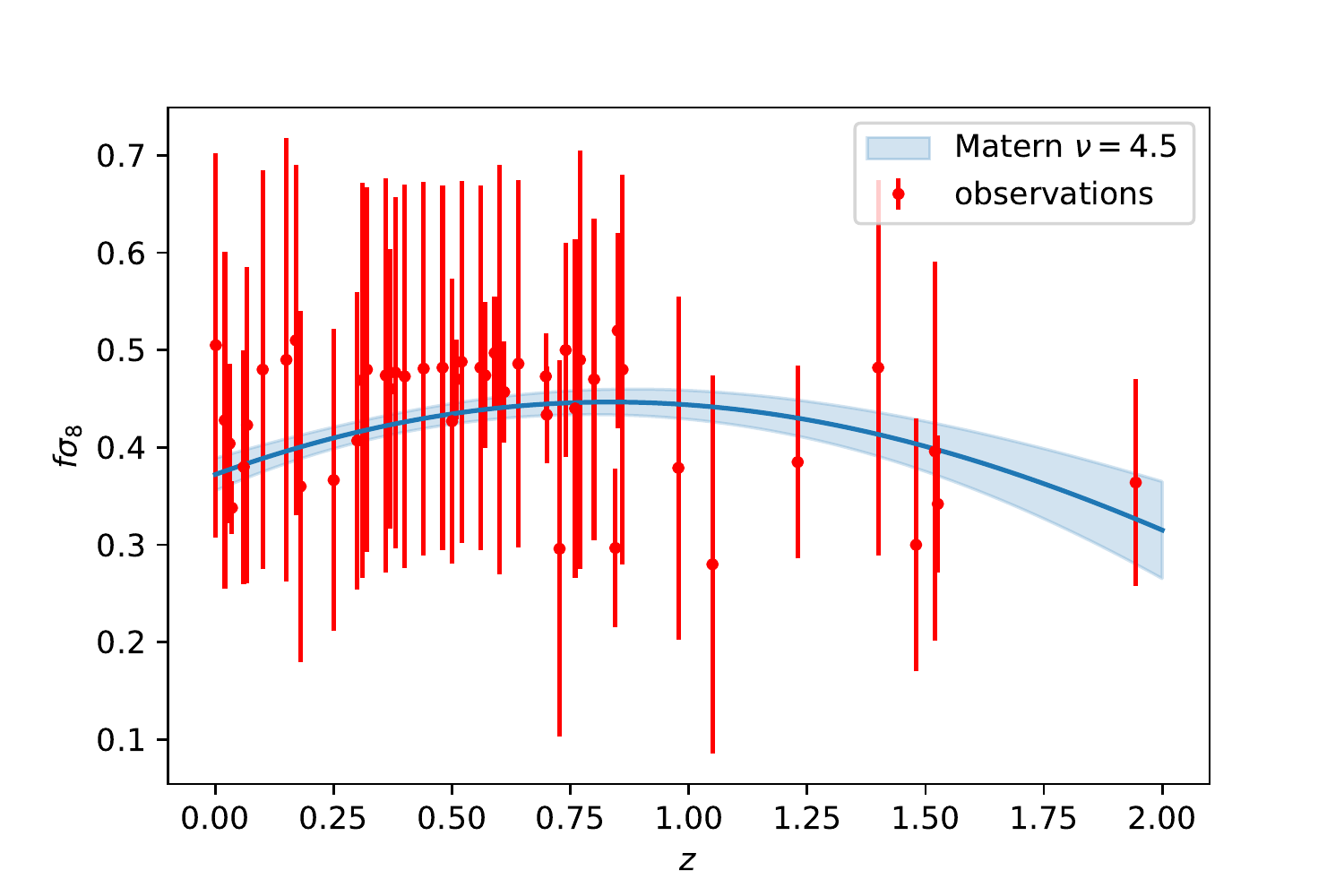}\\
 	\label{fig:3}
 	
\begin{tabular}{cccc} \hline \hline
 Kernel & RBF & Matern $\nu = 3.5$ & Matern $\nu = 4.5$
 \vspace{0.05cm}\\ \hline
\hline
$f\sigma_8(0)$ & $0.358\pm 0.0241$ & $0.356 \pm 0.025 $ & $0.372 \pm 0.017$ \\ 

\hline\hline
\end{tabular}
\caption{\it{The growth of matter data set with the corresponding data from the Gaussian Process Regression for different Kernels. The predicted shape presented with 1$\sigma$ error. The initial value of $f\sigma_8$ from the GP with different kernels. \label{tab:Results2}}}
\end{figure}

{Fig \ref{fig:2} presents the posterior distribution for $\Lambda$CDM and wCDM. The red distribution shows the fit for diagonal covariance and the blue distibution show the fit with random covariance between the point. $\Lambda$CDM fit yields $\Omega_m = 0.201^{+0.036}_{-0.033}   $,  $\sigma_8 = 0.857^{+0.044}_{-0.042}   $ and $S_8 = 0.700^{+0.038}_{-0.037} $. Including the random convariance matrix we get $\Omega_m = 0.192^{+0.037}_{-0.033} $,  $\sigma_8 =0.863^{+0.046}_{-0.043}$ and $S_8 = 0.690^{+0.040}_{-0.038} $ .}

{The wCDM model fit yields  $\Omega_m = 0.198^{+0.038}_{-0.035}$, $\sigma8 = 0.891^{+0.11}_{-0.095}$ and $S_8 = 0.722^{+0.072}_{-0.067}$. The equation of state in this case is: $w = -0.90^{+0.29}_{-0.30} $. Including the random convariance matrix we get $\Omega_m = 0.192^{+0.038}_{-0.034}$, $\sigma_8 = 0.879^{+0.10}_{-0.088} $ and $S_8 = 0.701^{+0.077}_{-0.074}$. The equation of state in this case is $w = -0.96^{+0.30}_{-0.31} $.}

{\cite{Omori:2018cid} measures the cross-correlation between red MaGiC galaxies selected from the DES Year-1 data and gravitational lensing of the CMB reconstructed from South Pole Telescope (SPT) and Planck data. Joint analysis of galaxy-CMB lensing cross-correlations and galaxy clustering to constrain cosmology finds $S_8 = 0.800^{+0.090}_{-0.094}$. The tension between the RSD fit to the DES fit is $0.98 \sigma$. Planck result gives $0.834 \pm 0.016$ \cite{Aghanim:2018eyx}. The difference with the Planck fit is $3.2\sigma$. The tension with the Planck data implies that the tension is real.}

\section{Gaussian Process Method}
\label{sec:GPM}
{Gaussian Process Method (GPM) is widely used in cosmology \cite{Seikel:2012uu,Bengaly:2019oxx,LHuillier:2019imn,Liao:2019qoc,Zhang:2018gjb,Gomez-Valent:2018hwc,Melia:2018tzi,Yang:2019fjt,Bengaly:2019ibu,Velasquez-Toribio:2019voz,Mehrabi:2020zau,Basilakos:2014yda,LHuillier:2017ani,Kase:2019mox,Liao_2019,Koo:2020ssl,Aljaf:2020eqh,Arjona:2020kco,Arjona:2020axn}. In this section we use the GPM with the $f\sigma_8$ data in order to estimate the $S_8$ with a model independent approach.} \cite{Adams:2017val,Beutler:2012px,Adams:2020dzw} estimate the value of $f\sigma_8 $ for $z \approx 0$. \cite{Adams:2017val} measures $0.424^{+0.067}_{-0.064}$ while \cite{Adams:2020dzw} measures $0.384 \pm 0.052$. In order to find the $f\sigma_8 (0)$ we use the Gaussian Process (PG) algorithm as a model independent approach. The GP reconstructs a function from data set without assuming a parametrization for the function \cite{Seikel:2012uu,Lyu:2019rzl}. Having a data set $D$:
\begin{equation}\label{eq:data-set}
	D=\{(x_i,y_i)|i=1,..,n\},
\end{equation}
we can reconstruct in a function $f(x)$ which describes the data. In this case at any point $x$, the value $f(x)$ is a Gaussian random variable with mean $\mu(x)$ and variance $Var(x)$. The function values at any two different points are not independent from each other. Therefore, the covariance function $cov(f(x),f(\tilde{x}))=k(x,\tilde{x})$ 
describes the corresponding correlations. The possibilities for the Kernel are wide. The current work uses the Radial Basis Function (RBF):
\begin{equation}\label{eq:cov-squ}
k(x,\tilde{x}) = \sigma_f^2\exp(-\frac{(x-\tilde{x})^2}{2l^2}),
\end{equation}
The Matern kernel with $\nu = 7/2$:
\begin{equation}\label{eq:cov-m72}
\begin{split}
k(x,\tilde{x}) =  \sigma_f^2\exp(-\sqrt{7}\frac{|x-\tilde{x}|}{l})\\(1+\sqrt{7}\frac{|x-\tilde{x}|}{l}+14\frac{(x-\tilde{x})^2}{5l^2}+7\sqrt{7}\frac{|x-\tilde{x}|^3}{15l^3}),   
\end{split}
\end{equation}  
and Matern kernel with $\nu = 9/2$:
\begin{equation}\label{eq:cov-m92}
\begin{split}
  k(x,\tilde{x}) =  \sigma_f^2\exp(-3\frac{|x-\tilde{x}|}{l})\\(1+3\frac{|x-\tilde{x}|}{l}+27\frac{(x-\tilde{x})^2}{7l^2}+18\frac{|x-\tilde{x}|^3}{7l^3}+ 27\frac{(x-\tilde{x})^4}{35l^4}).
\end{split}
\end{equation} 
$\sigma_f$ and $l$ are two hyperparameters 
which can be constrained from the observational data. In order to calculate the predicted behavior from the Gaussian Process method, we use the open source code \textbf{Scikit-learn} \cite{scikit-learn}.

{Fig \ref{fig:3} shows the smooth behavior for different Kernels. The table below presents the corresponding $f\sigma_8 (z=0)$. The RBG kernel yields $0.358\pm 0.0241$ while the Matern Kernel gives $0.356 \pm 0.025 $ for $\nu = 3.5$ and $0.372 \pm 0.017$ for $\nu = 4.5$. The total estimation gives closer value to \cite{Adams:2020dzw}'s measurement.}

\begin{figure}[t!]
 	\centering
\includegraphics[width=0.47\textwidth]{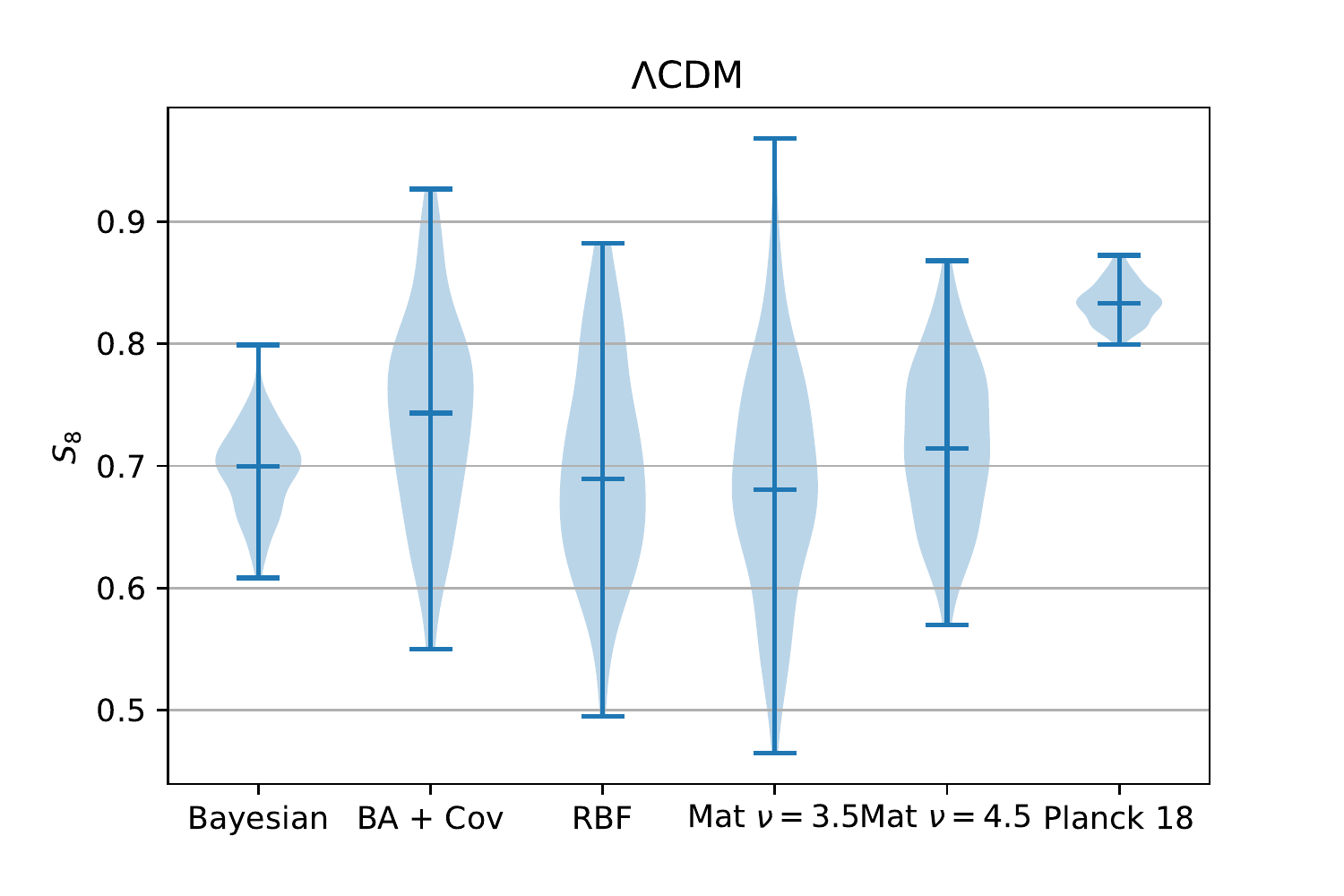}
\includegraphics[width=0.47\textwidth]{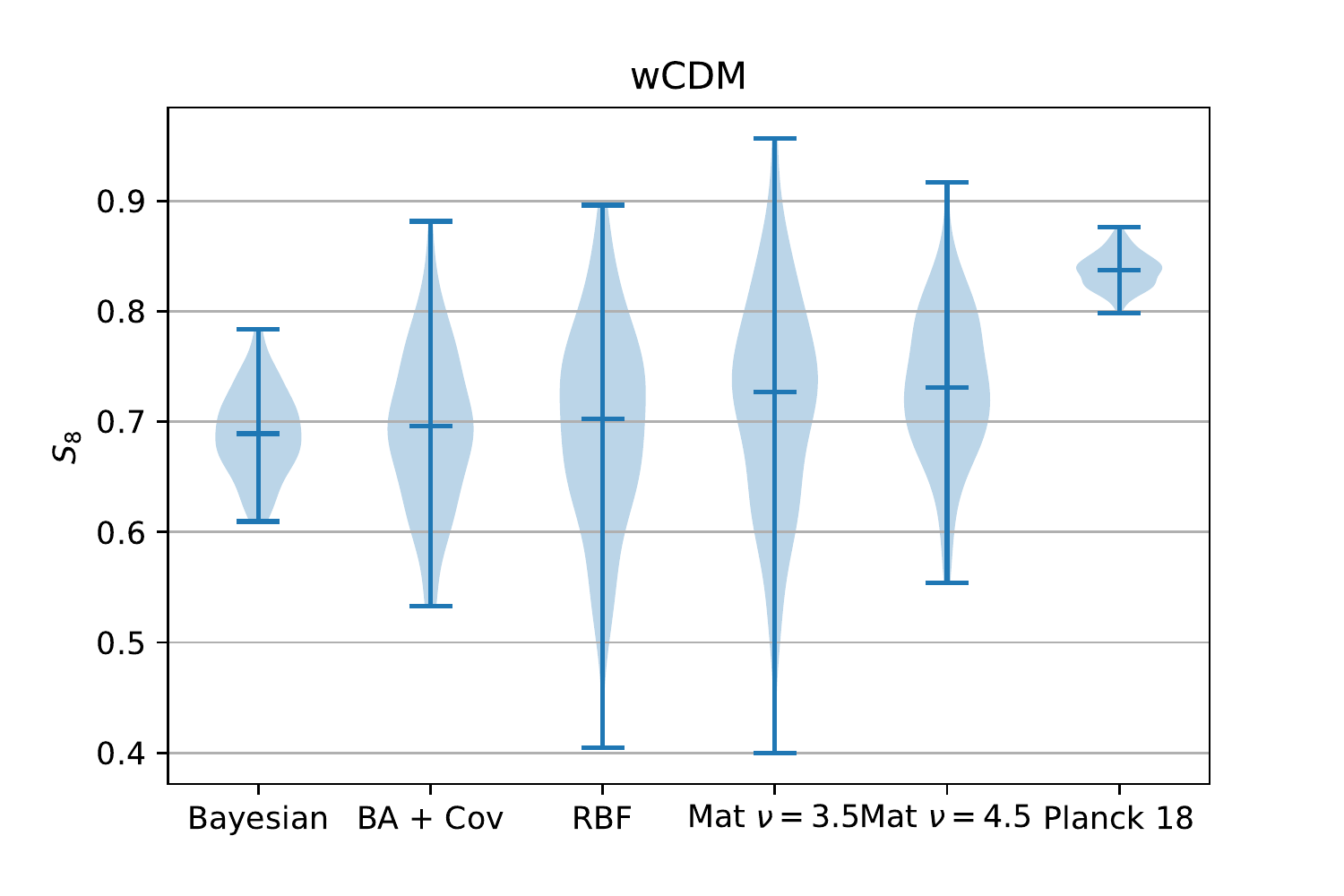}
\\

\begin{tabular}{cccc} \hline \hline
 $\Lambda$CDM & RBF & Matern $\nu = 3.5$ & Matern $\nu = 4.5$
 \vspace{0.05cm}\\ \hline
\hline
$\sigma_8$ & $0.77^{+0.21}_{-0.16}      $ & $0.77^{+0.21}_{-0.16}      $ & $0.79^{+0.19}_{-0.17}      $ \\ 
$\Omega_m$ & $0.27^{+0.12}_{-0.11}      $ &  $0.26^{+0.13}_{-0.11}      $ & $0.27^{+0.12}_{-0.10}      $
\\
$S_8$ &  $0.707^{+0.085}_{-0.085}   $ & $0.701^{+0.089}_{-0.089}   $ &  $0.731^{+0.063}_{-0.062}   $
 \\ 
\hline\hline
 wCDM & RBF & Matern $\nu = 3.5$ & Matern $\nu = 4.5$
 \vspace{0.05cm}\\ \hline
\hline
w & $-1.01^{+0.48}_{-0.47}     $ & $-0.998^{+0.47}_{-0.47}    $ & $-1.00^{+0.47}_{-0.48}     $
\\
$\sigma_8$ & $0.78^{+0.20}_{-0.17}      $ & $0.78^{+0.20}_{-0.17}      $ & $0.77^{+0.20}_{-0.16}      $ \\ 
$\Omega_m$ & $0.26^{+0.12}_{-0.11}      $ & $0.26^{+0.12}_{-0.11}      $ & $0.28^{+0.11}_{-0.11}      $ \\ 
$S_8$ & $0.704^{+0.087}_{-0.082}   $ & $0.706^{+0.098}_{-0.092}   $ & $0.730^{+0.067}_{-0.063}   $
\\
\hline\hline

\end{tabular}
 	\label{fig:4}
 \caption{{\it{The final results of $S_8$ estimations from different methods: the Bayesian analysis and the Gaussian Process Regression estimation. The models that were tested are $\Lambda$CDM (upper panel) and wCDM models (lower panel).}}}
\end{figure}

{From the dependence of Eq.~\ref{eq:delta} and Eq.~\ref{eq:fsig8z} we can estimate the parameters with the predicted point $f\sigma_8(0)$. The table under figure \ref{fig:4} summarizes the best fitted values with different kernels.}

{The GPM $\Lambda$CDM fit yields very similar values for the $S_8$. The RBF kernel gives: $S_8 = 0.707^{+0.085}_{-0.085}   $. The Matern kernel gives $S_8 = 0.701^{+0.089}_{-0.089}   $ for $\nu = 3.5$, and   $S_8 = 0.731^{+0.063}_{-0.062}   $ for $\nu = 4.5$. The wCDM fit gives very similar values: $S_8 = 0.704^{+0.087}_{-0.082}$. The Matern kernel gives $S_8 = 0.706^{+0.098}_{-0.092}$ for $\nu = 3.5$, and   $S_8 = 0.730^{+0.067}_{-0.063} $ for $\nu = 4.5$.}

{With the GPM the tension between the Planck $S_8$ fit and the RSD fit is reduced: $1.49\sigma$ for the RBF kernel, $1.47\sigma$ for the Matern kernel with $\nu = 3.5$ and $1.58\sigma$ for the Matern kernel with $\nu = 4.5$. The results imply that the tension is probably real and gives some window for a new physics, however the tension most probably small. The possibility for new physics is widely discussed with different models \cite{Saridakis:2019qwt,DiValentino:2020srs,DiValentino:2020vvd,DiValentino:2020zio,DiValentino:2020vhf,Abadi:2020hbr}.} 

{In order to complete our analysis, we report the tension for the wCDM model: $1.47\sigma$ for the RBF kernel, $1.29\sigma$ for the Matern kernel with $\nu = 3.5$ and $1.51\sigma$ for the Matern kernel with $\nu = 4.5$.}

\section{Discussion}
\label{sec:Dis}

{This paper analyzes the latest $f\sigma_8$ data with the standard Bayesian statistics and model independent approach. From a big collection of data points we find a sub-collection for different redshifts. With  Bayesian statistics we find $S_8 = 0.700^{+0.038}_{-0.037}$ for $\Lambda$CDM and $S_8 = 0.722^{+0.072}_{-0.067} $ for wCDM. }

{With a model independent approach, we find the initial value $f\sigma_8 (z=0) \approx 0.36 \pm 0.02$. A fit with this initial point we get closer values of $S_8$ to the Planck value: In all cases, the tension between Planck 2018 fit and the RSD fit is smaller then $2 \sigma$. In future measurements the tension may be resolved, such as J-PAS \cite{Benitez:2014ibt}, DESI \cite{Aghamousa:2016zmz} and Euclid experiment \cite{Laureijs:2011gra,Sprenger:2018tdb,Tutusaus:2020xmc} or at least reduced as we see in our analysis, but the possibility the tension is real is plausible solution. }

\textbf{Public Source}: The python files with the dataset and the fit package can be found in \url{https://github.com/benidav/RSD-GPM}{}.

\acknowledgements
I would like to thank E. Kovetz, E .I. Guendelman, D. Vasak and D. Staicova for discussions and advice. I gratefully acknowledge the support from Frankfurt Institute for Advanced Studies (FIAS) as well the support from Ben-Gurion University in Beer-Sheva, Israel. I have received partial support from European COST actions CA15117 and CA18108. I am also thankful to Bulgarian National Science Fund for support via research grant KP-06-N 8/11.

\bibliographystyle{apsrev4-1}
\bibliography{ref}

\end{document}